\documentclass[10pt, conference, compsocconf]{IEEEtran}

\usepackage{graphicx}
\usepackage{graphics}
\usepackage{framed}
\usepackage{subfig}
\usepackage{longtable}
\usepackage{multirow}
\usepackage{hyperref}
\usepackage{amsmath}
\usepackage{algorithm}
\usepackage[noend]{algpseudocode}

\makeatletter
\def\BState{\State\hskip-\ALG@thistlm}
\makeatother

\ifCLASSINFOpdf
\else
\fi
\usepackage{array}
\usepackage{url}

\begin{document}
%
\title{Affinity Prediction in Online Social Networks}



%
\author{\IEEEauthorblockN{Matias Estrada\IEEEauthorrefmark{1}\IEEEauthorrefmark{2} and
Marcelo Mendoza\IEEEauthorrefmark{2}}
\IEEEauthorblockA{\IEEEauthorrefmark{1}Skout Inc., Chile}
\IEEEauthorblockA{\IEEEauthorrefmark{2}Universidad T\'ecnica Federico Santa Mar\'ia, Chile}}


\maketitle

\begin{abstract}
Link prediction is the problem of inferring whether potential edges between pairs of vertices 
in a graph will be present or absent in the near future. To perform this task it is usual to use information provided by a number of available and observed vertices/edges. Then, a number of edge scoring methods based on this information can be created. Usually, these methods assess local structures of the observed graph, assuming that closer vertices in the original period of observation will be more likely to form a link in the future. In this paper we explore the combination of local and global features to conduct link prediction in online social networks. The contributions of the paper are twofold: a) We evaluate a number of strategies that combines global and local features tackling the locality assumption of link prediction scoring methods, and b) We only use network topology-based features, avoiding the inclusion of informational or transactional based features that involve heavy computational costs in the methods. We evaluate our proposal using real-world data provided by Skout Inc.~\footnote{\url{http://www.skout.com}}, an affinity online social network with millions of users around the world. Our results show that our proposal is feasible.
\end{abstract}


%
\IEEEpeerreviewmaketitle
\vspace{-2mm}
\section{Introduction}
\label{sec:intro}

An online social network is the ideal place for meeting new people. 
A number of web platforms supporting the emergence of new communities offer services 
connecting people and making possible a wide range of social connections. 
The ability to opportunely detect a potential social connection between two users is a key factor for the growth and success of these networks. 

Link prediction is the problem of inferring whether potential edges between pairs of vertices 
in a graph will be present or absent in the near future. 
In a link prediction task, we first assume the existence of an original observed graph, with a completely known set of vertices and a partially known set of edges. 
Accordingly, the link prediction task is to infer the rest of the edges.

Let $G = (V,E)$ be an undirected graph. At a given point in time $t_0$, we assume that all vertices $v \in V$ are known but only a subset of $E$ is known. Let $E^{obs}$ be the subset of edges $e \in E$ that is known at $t_0$, and let $E^{miss}$ be the subset of unobserved edges at $t_0$, such that $E^{miss} = E \setminus E^{obs}$. The problem of link prediction is to predict the edges in $E^{miss}$ from $G = (V,E^{obs})$. 

Usually, $G$ is sparse, i.e. $\mid E \mid \ll \mid V \times V \mid$. A commonly used approach for link prediction is to exploit locality in $G = (V,E^{obs})$. The basic assumption is that is more likely to observe an edge $e=(u,v)$ in $E^{miss}$ if $u$ and $v$ are closer vertices in $G = (V,E^{obs})$. In particular, link prediction methods use network locality.

A natural way to explore in link prediction methods is to combine local features and global features. 
The combination of global and local features has offered several benefits to information retrieval problems, such as document ranking, where local features (i.e. query dependent features such as cosine distance) and global features (i.e. query independent features such as PageRank) are combined in ranking functions improving precision. Following this approach, we will explore the use of local/global features for link prediction. To do this, we will separate the link prediction problem into two tasks:  1) A retrieval step where a number of potential edges for a given query $q$ are retrieved, and 2) A ranking step where the potential edges are sorted by relevance to $q$.

Our proposal considers local features for the retrieval of potential edges. 
However, for our approach the locality constraint is considered as a soft constraint. 
We do this by introducing a threshold for locality, which allow us to control the length of the list of potential edges. 
Then, the list of candidates is sorted by considering global features, such as authority or transitivity measures. 

The use of local/global features introduce a precision/recall tradeoff. 
The lower the locality threshold, the better the recall. 
The opposite is also true. 
The higher the locality threshold, the better the precision. 
The search for a balance between the use of global/local features is necessary to deal with 
this tradeoff.

We run our methods over real world data provided by Skout Inc. 
Skout is an affinity online social network with millions of users around the world. 
A full subgraph of Skout was procesed, retrieving an initial graph with almost 2 million of users and 4 million of edges between them. 
Then, during one month, the evolution of this subgraph was tracked, registering a densification that reaches almost 600,000 new edges. 
We explore the feasibility of our proposal trying to predict edges that were created during the period of observation. 
Our results show that the proposal is feasible. 

This article is organized as follows. 
In Section~\ref{sec:relwork} we discuss related work. 
We introduce our affinity prediction strategy in Section~\ref{sec:strategy}. 
In Section~\ref{sec:features} we discuss about the features used by our trategy. 
Experimental results are presented in Section V. 
Finally, we conclude in Section~\ref{sec:conclusions} with a brief discussion about some open questions and future work. 

\section{Related Work}
\label{sec:relwork}

User recomendation is mainly achieved through the transitivity properties~\cite{newman2001} or friend-of-a-friend~\cite{silva2010}. If a user A is friend of a user B and the user B is friend of a user C, then A and C probably will be friends~\cite{kossinets2006}. In general, online social networks are sparse~\cite{delgenio2011}, meaning that the total amount of potential links to be created is much greater than the links that are actually created. Sparsity turns the link prediction problem into a very dificult problem returning success improvements over a random predictor from 3\% to 54\%~\cite{liben2003}.

In the last years a number of different algorithms has been developed to predict new links on many types of networks like scientific collaboration networks~\cite{liben2003}, protein networks~\cite{lu2009}, energy grid networks~\cite{lu2009} and social networks~\cite{murata2008}, among others.

A number of different types of similarity indexes is available for this problem, among them user-based or vertex-based indexes~\cite{lin1998}. Other similarity indexes widely used are based on local or global network topology measures. Locality measures are based on the user's subgraph and do not require the full graph from being stored. One of the simplest indexes is the Common Neighboors \cite{newman2001}. This index considers the amount of common neighboors that two nodes have. When the number of common neighboors is higher, then is more likely that those nodes create a link in the future~\cite{kossinets2006}. The Jaccard index~\cite{jaccard1912} compares the common neighboors cardinality with the cardinality of the union of both neighboors, getting a value that represents the proportion between the cardinality of these two sets. Other indexes often used are the Salton index~\cite{salton1983}, also known as cosine similarity, and the Srensen index~\cite{sorensen1948} that is mainly used in ecological networks~\cite{linyuan2010}. Other well known locality measure is the Adamic \& Adar index ~\cite{ada2003}.

Global indexes require the entire network to be examined. One of these indexes is known as the Katz index~\cite{katz1953}. The Katz index adds paths dumped by a dumping factor. If this factor is small, then the index will behave similar to the Common Neighbours index. The Local Path index~\cite{lu2009} is similar to the index proposed by Katz, but it uses a limited path lenght requiring less computational complexity. A recursive index known as SimRank index~\cite{jeh2002} is calculated using a random walk process, that propagated through the graph with a decay factor. 

\section{An Affinity Prediction Strategy}
\label{sec:strategy}

\begin{figure}[h!]
\begin{center}
    \fbox{\includegraphics[width=0.9\columnwidth]{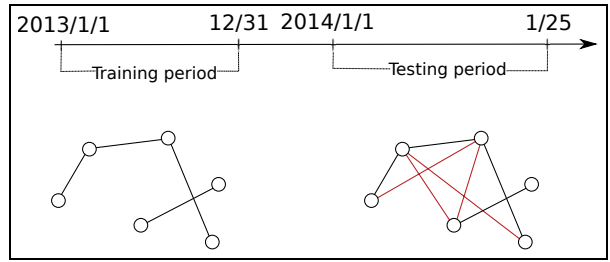}}
  \caption{Link prediction as a link classification problem}
  \label{fig-0}
\end{center}
\end{figure}

Let $G = (V,E)$ be an undirected graph. At a given point in time $t_0$, we assume that all vertices $v \in V$ are known but only a subset of $E$ is known. Let $E^{obs}$ be the subset of edges $e \in E$ that is known at $t_0$, and let $E^{miss}$ be the subset of unobserved edges at $t_0$, such that $E^{miss} = E \setminus E^{obs}$. As the problem of link prediction is to predict the edges in $E^{miss}$ from $G = (V,E^{obs})$, a natural way to address the problem is to create a model from $G = (V,E^{obs})$ capable of predicting potential edges in $E^{miss}$ using:

\[
\mathcal{P} \left( E^{miss} \mid E^{obs}, \mathcal{X}_v \right) ,
\]

\noindent where $\mathcal{X}_v$ is a collection of vertex features retrieved from $G$. 
It is usual to estimate a model $\hat{\mathcal{P}}$ by using labeled data instances over $\mathcal{X}_v$, modeling the link prediction task as a real/false link classification problem. 
To do this, a set of edges $E_0$ is observed in a period subsequent to $t_0$, assuming that these edges are good descriptors of edges in $E^{miss}$. Then, these edges are characterized over $\mathcal{X}_v$ and labeled as real link examples. From the set $E \times E \setminus \{ E_0 \cup E^{obs} \}$ a random sample of edges is retrieved, characterized over $\mathcal{X}_v$ and labeled as false link examples. Thus, a solver is trained to build $\hat{\mathcal{P}}$ according to a given criterion function. Figure~\ref{fig-0} illustrates the approach described above. 

It is usual to take into consideration a long period of time to observe the graph and calculate $\mathcal{X}_v$. 
In the machine learning community this period is known as training period. 
In Figure~\ref{fig-0} the training period ranges a whole year and $t_0$ is the last day of that year (december 31). After $t_0$ and during 25 days, new links (depicted by red lines) are registered and stored to create a collection of real links. This period of time is named teting period. Every link not observed during the training and testing periods is considered as a false link. Then, false links (every link not included in the figure) and real links (red arcs) are used to create a data set with real/false examples characterized over $\mathcal{X}_v$.

A number of factors can help explain the success of a such strategy. 
First, the success depends on the quality of the machine learning solver. 
It is usual to deal with overfitted models limiting the capability of $\hat{\mathcal{P}}$ to generalise to new data instances. 
The creation of a model with good generalisation properties is far to be an easy task. 
In addition, the success depends on the choice of $t_0$ and on the existence of a considerable training and testing period. 

We want to compare the strategy described above to a ranking-based strategy. 
A ranking-based strategy considers two steps to rank links. For a given vertex $u$, a vertex retrieval 
step is conducted, using a locality constraint to recover a list of proximal vertices to $u$. 
Then, the list is sorted according to a given relevance criteria. 
The following pseudocode illustrates the ranking-based strategy.

\begin{algorithm}
\caption*{Ranking-based Link Prediction}\label{algo}
\begin{algorithmic}[1]
\Procedure{Link Ranking}{$u,th$}
\BState \emph{retrieval}:
\State $\mbox{seeds} \gets ()$
\State $\Gamma (u) \gets \text{fetch neighbor list of }u$
\For{\textbf{each} v in $\Gamma (u)$}
\If {$\mbox{locality}(u,v) > th$} 
\State $\mbox{seeds} \gets v$
\EndIf
\EndFor
\BState \emph{ranking}:
\State $\mbox{scores} \gets ()$
\For{\textbf{each} v in $\mbox{seeds}$}
\For{\textbf{each} c in $\Gamma (v)$}
\State $\mbox{scores[c]} \gets \mathcal{X}_v(c)$
\EndFor
\EndFor
\State \Return $\mbox{top K elements of scores}$
\EndProcedure
\end{algorithmic}
\end{algorithm}

As the pseudocode indicates, two steps are required to address the link prediction task as a ranking-based problem. The first step performs a retrieval step over the set of neighbors of $u$. To do this, a localitity threshold constraint is verified, putting these vertices into a list of seeds. Then, in the ranking step of the algorithm, for each neighbor $c$ of each seed retrieved in the previous step, a score is calculated over $\mathcal{X}_v(c)$. Finally, the top K elements of the list of candidates are returned.

\section{Global - Local Features for Affinity Prediction}
\label{sec:features}

\subsection{Local Features}

We will explore the use of one locality measure: Jaccard 
The definition of this measures is given below.

\paragraph{Jaccard coefficient} Let $G = (V,E^{obs})$ be the observed graph at $t_0$ time. 
Let $u$ and $v$ be a pair of vertices in $V$, and let $\Gamma (u)$ and $\Gamma (v)$ be the neighborhood of $u$ and $v$, respectively. The Jaccard similarity coefficient of $u$ and $v$ is the proportion between common neighbors and the union of both neighborhood. Formally:

\[
\tt{Jaccard}(u,v) = \frac{\mid \Gamma (u) \cap \Gamma (v) \mid}{\mid \Gamma (u) \cup \Gamma (v) \mid}.
\]

\subsection{Global Features}

We will explore the use of three global measures for link prediction: Normalized degree, HITS-based measures, and transitivity. These features are query independent, that is to say, these measures define a collection of point wise estimations at vertex level. The value of a measure of this collection is the same for the whole graph, therefore we call these measures global features. The definitions of these measures is given below.

\paragraph{Degree coefficient} Let $\Gamma (u)$ be the neighborhood of a vertex $u$ and let $\mid \Gamma (u) \mid$ be the cardinality of this set, also known as the degree of $u$. We define a degree coefficient by normalizing $\mid \Gamma (u) \mid$ with the maximum degree of $G$. Formally:

\[
\tt{Degree}(u) = \frac{\mid \Gamma (u) \mid}{\tt{Max}_{v \in G} \mid \Gamma (v) \mid}.
\]

\paragraph{Transitivity coefficient} Let $\Gamma (u)$ be the neighborhood of a vertex $u$. The transitivity coefficient $\tt{Transitivity}(u)$ (a.k.a. clustering coefficient) is the ratio between the number of links in $\Gamma (u)$ and the maximum number of such links. If $\Gamma (u)$ 
has $e_u$ links, we have:

\[
\tt{Transitivity}(u) = \frac{e_u}{\frac{\mid \Gamma (u) \mid \cdot (\mid \Gamma (u) \mid - 1)}{2}} .
\]

\paragraph{HITS coefficients} HITS (Hypertext Induced Topic Search) coefficients come from the information retrieval community, were proposed by Kleinberg \cite{kleinberg} and were firstly used for web page ranking. The idea is that pages that have many links pointing to them are called authorities and pages that have many outgoing links are called hubs. Good hubs point to good authority pages, and vice versa. Lets $\tt{hub}(u)$ and $\tt{auth}(u)$ be hub and authority coefficients for a vertex $u$. The following equations can be solved through an iterative algorithm that addresses the fixed point problem defined by:

\[
\tt{hub(u) = \sum_{v \in G \mid u \rightarrow v} \tt{auth}(v)} , 
\]

\[
\tt{auth(u) = \sum_{v \in G \mid v \rightarrow u} \tt{hub}(v)} .
\]

In undirected graphs, both coefficients have the same value. 

\begin{figure}[h!]
\begin{center}
\includegraphics[width=0.6\columnwidth]{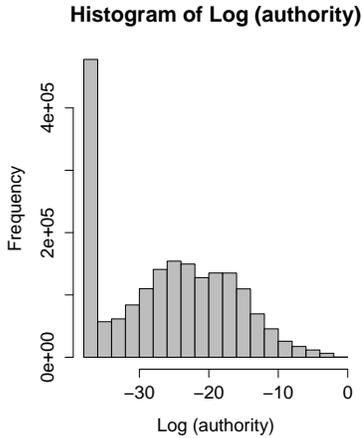}
\end{center}
\caption{Histogram of Authority Scores.}
\label{fig:auth}
\end{figure}

\begin{figure}[h!]
\begin{center}
\includegraphics[width=0.6\columnwidth]{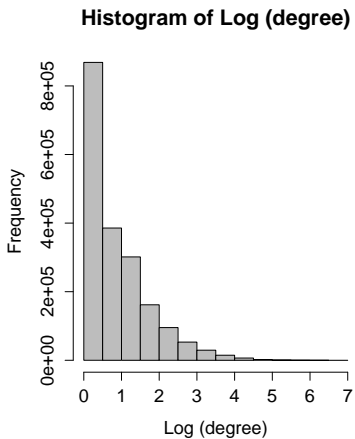}
\end{center}
\caption{Histogram of Degree Scores.}
\label{fig:degree}
\end{figure}

\begin{figure}[h!]
\begin{center}
\includegraphics[width=0.6\columnwidth]{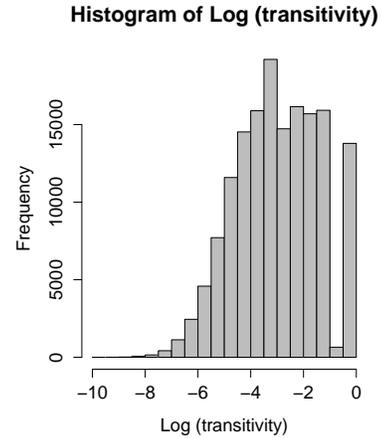}
\end{center}
\caption{Histogram of Transitivity Scores}
\label{fig:trans}
\end{figure}

We retrieved the link graph of Skout that corresponds to a whole year (2013).  
For the collection of 542,010 users that created at least one new link during the observation period, 
we calculated the three measures introduced above. These histograms are in Figures~\ref{fig:auth}, ~\ref{fig:degree} and ~\ref{fig:trans}.

Figures~\ref{fig:auth}, ~\ref{fig:degree} and ~\ref{fig:trans} shows histograms for the log(A+1) measure of authority, degree and transitivity. 
Authority shows a significant proportion of users in the first bin. However a number of users is concentrated in the range between -30 and -10. Something similar occurs in the histogram of log transitivity, where the coefficientes are concentrated around -4 and a significant bin in cero indicates the presence of cliques. Finally, we can observe that the normalized degree feature is concentrated around low values. 
 
\section{Experimental Evaluation}

\subsection{Results}
\label{sec:results}

We explore the link prediction task using real world data retrieved from Skout. 
We registered every link created during one year (2013) achieving an unidirected graph with 3,855,389 links between 1,920,015 users. 

\begin{figure}[h!]
\begin{center}
\includegraphics[width=0.9\columnwidth]{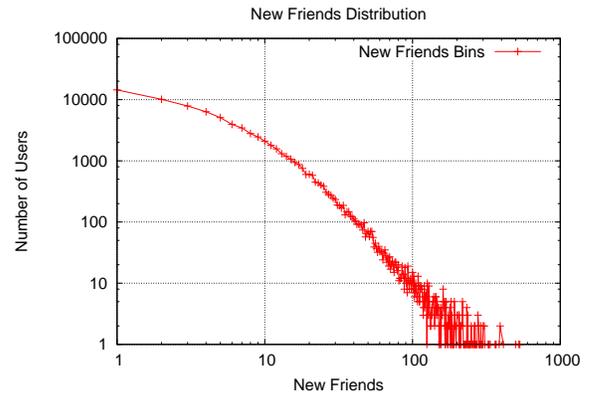}
\end{center}
\caption{Number of links created per users during the observation period.}
\label{fig-2}
\end{figure}

Then we registered the links created during the first 25 days of January 2014. Our collection of links registered 582,199 new links created during this period between users with accounts created before the 1st of January (old users from now and so on). A total of 428,341 old users achieved a new friend during the observation period. The distribution of new friends per user is shown in Figure~\ref{fig-2}.

As Figure~\ref{fig-2} shows, the distributions of new friends per user follows a rich get richer law: Only a few users concentrates a lot of new friends whilst the majority of the users achieved only one or two new friends during the period. 

\begin{figure}[h!]
\begin{center}
\includegraphics[width=0.9\columnwidth]{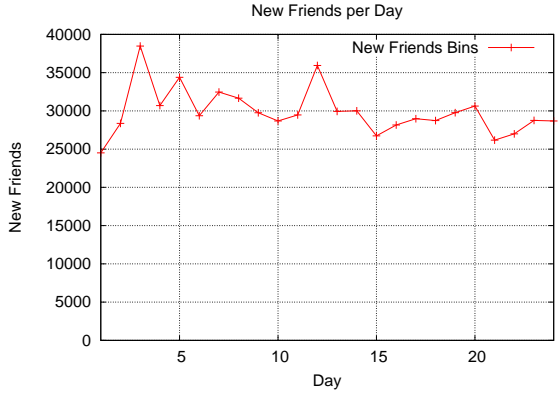}
\end{center}
\caption{Number of links created per day during the observation period.}
\label{fig-3}
\end{figure}

The number of links created per day during the observation period is showed in Figure~\ref{fig-3}. 
We can observe there that only two days exhibit peaks in the creation of new links (the third and twelfth day) and the other ones are concentrated around 30,000 links per day.

\begin{figure}[h!]
\begin{center}
    \includegraphics[width=0.8\columnwidth]{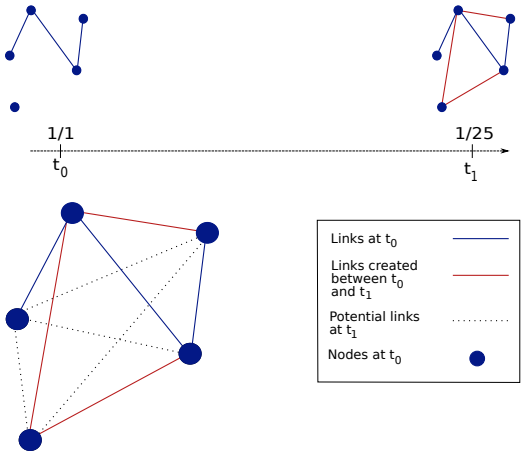}
  \caption{Data set creation methodology. Blue links and nodes depict the graph at $t_0$. Red links depict links created during the observation period. Links depicted with dashed lines indicates potential unobserved links. In our strategy, dashed links are labeled as false link instances and red links as real link instances.}
  \label{fig-meth}
\end{center}
\end{figure}

A total amount of 76,848 users accepted at least one new friend during the observation period. 
We named these users the "active users" of the period. 
The collection of potential links used to create our machine learning data set is the set of unobserved links that start from active users and point to users with at least one new link during the observation period, discarding from this data set potential links that may be created between inactive users. 
Then we reduced the classification problem to the separation between unobserved links and real links that start from active users. This methodology is described in Figure~\ref{fig-meth}.

This scenario encourages the use of machine learning techniques. 
As we use the information about which users created links during the observation period, 
the data set takes advantages of this information, reducing the link prediction problem to 
a classification between real/false links between active users. In fact, the real world problem is much more difficult, because we do not have apriori information about active users. Therefore, the classifier 
has to deal with the whole set of potential links, that exponentially growths with $V$. Despite these considerations, this approach is useful to illustrate how global measures behave in this problem, uncovering some data properties. 

We balanced real and false link data instances in our data set to avoid biased results. 
We do this by sampling uniformly at random the whole set of labeled data instances. 
As a result we achieved a data set with 343,887 real link labeled data instances, and 345,000 false link labeled data instances. 

For each data instance considered in our data set and for each vertex of the labeled link, we calculated authority, normalized degree and transitivity measures. We show in Table~\ref{tab:inf} information gain values for each feature considered in our data set.

\begin{table}[h!]
\begin{center}
\begin{tabular}{|l|c|}
\hline 
Feature& Information Gain\\
\hline \hline
Authority 2& 0.9155\\
\hline 
Authority 1& 0.5367\\
\hline 
Transitivity 2& 0.1192\\
\hline 
Degree 2& 0.1054\\
\hline 
Transitivity 1& 0.0392\\
\hline 
Degree 1& 0.0227\\
\hline 
\end{tabular}
\end{center}
\caption{Information Gain values for each feature considered in our data set. Active users features are depicted with a subindex equals to one.}
\label{tab:inf}
\end{table}

As Table~\ref{tab:inf} shows, the most relevant feature for this problem is the authority score of the second vertex. As the first vertex corresponds to the active user (the one that accepts/rejects the friendship request), the authority of the second user is a measure of the visibility of the candidate user for the rest of the graph (how strongly connected is the candidate user to the graph). On the other hand, features of the active user that are related to its locality are only marginally relevant for the problem, indicating that the current connectivity of the active user neighborhood does not correlate to the creation of new links.   

\begin{figure}[h!]
\begin{center}
    \fbox{\includegraphics[width=0.9\columnwidth]{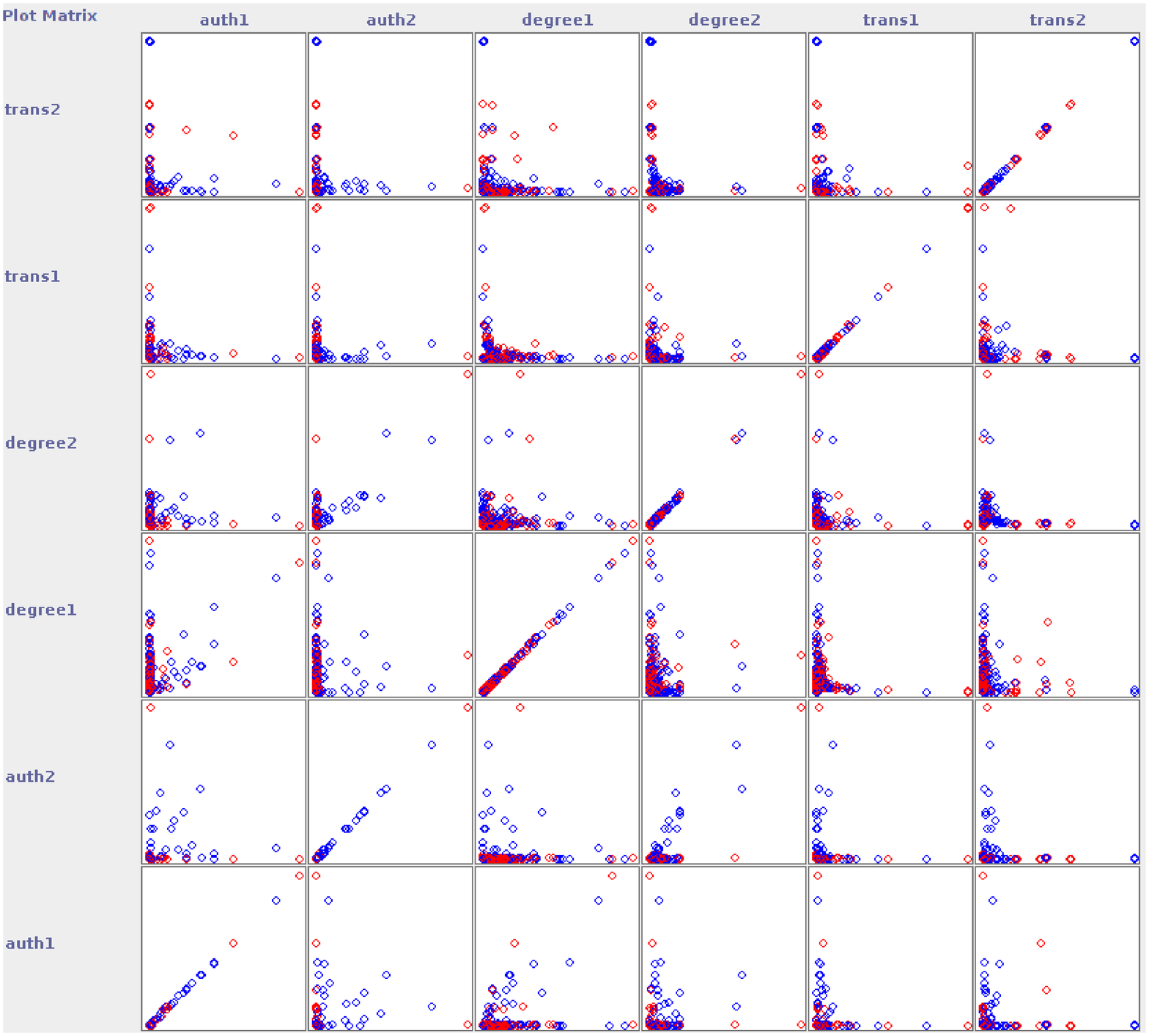}}
  \caption{Correlation Matrix for the features considered in our data set. Real links are depicted with red points and false links with blue points.}
  \label{fig-5}
\end{center}
\end{figure}

The plot matrix of the data set is depicted in Figure~\ref{fig-5}. 
We applied a logarithmic function to each feature to facilitate the visualization of the plots. 

As Figure~\ref{fig-5} shows, each pairwise feature comparison illustrates the absence of correlation. 
Accordingly, we discard the use of subset feature selection. 

In Table~\ref{tab:machine} we show performance results for the classifiers created to solve the real/false link classification problem. We used a 5-fold cross validation strategy to evaluate each classifier. The solvers used were Naive Bayes, J48 (decision trees) and Logistic Regression.

\begin{table*}[h!]
\begin{center}
\vspace{1cm}
{\setlength{\extrarowheight}{1.0pt}
\begin{tabular}{|l|l|c|c|c|c|c|c|c|c|}
\hline
& Class& TP Rate& FP Rate& Precision& Recall& F-Measure& MCC& ROC Area& PRC Area\\
\hline \hline

\multirow{3}{*}{Bayes} & real& 0.138& 0.035& 0.796& 0.138& 0.235& 0.183& 0.605& 0.635\\
\cline{2-10}  
& false& 0.965& 0.862& 0.529& 0.965& 0.683& 0.183& 0.605& 0.572\\
\cline{2-10} 
& Weighted Avg.& 0.552& 0.449& 0.662& 0.552& 0.460& 0.183& 0.605& 0.604\\
\hline \hline

\multirow{3}{*}{J48} & real& 0.642& 0.232& 0.734& 0.642& 0.685& 0.414& 0.789& 0.813\\
\cline{2-10} 
& false& 0.768& 0.358& 0.683& 0.768& 0.723& 0.414& 0.789& 0.778\\
\cline{2-10} 
& Weighted Avg.& 0.705& 0.295& 0.708& 0.705& 0.704& 0.414& 0.789& 0.795\\
\hline \hline

\multirow{3}{*}{Log} & real& 0.316& 0.132& 0.705& 0.316& 0.437& 0.221& 0.662& 0.666\\
\cline{2-10} 
& false& 0.868& 0.684& 0.560& 0.868& 0.681& 0.221& 0.662& 0.625\\
\cline{2-10} 
& Weighted Avg.& 0.593& 0.408& 0.633& 0.593& 0.559& 0.221& 0.662& 0.646\\
\hline 
\end{tabular}}
\end{center}
\caption{Performance measures for the real/false link classification problem.}
\label{tab:machine}
\end{table*}

As Table~\ref{tab:machine} shows, the best results are achieved by using a J48 decision tree, with almost a perfect performance balance between both classes. The other solvers exhibit biased results, despite the fact that the data set is label balanced. In fact, Naive Bayes and Logistic Regression biases their results to the detection of false links, with a high TP rate for this class. On the other hand, the results for real links are very poor, with TP rates equals to 0.138 and 0.316 for Bayes and Log Regression, respectively. These results imply low recall rates and, accordingly, low F-measure rates. Thus, we will analize the results achieved by using J48, discarding Bayes and Log Regression for this part of the data analysis.

\begin{figure}[h]
\begin{center}
    \fbox{\includegraphics[width=\columnwidth]{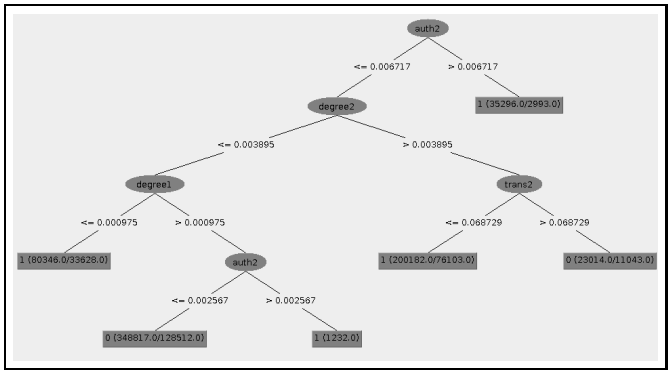}}
  \caption{Decision tree for the real/false link classification problem.}
  \label{fig-6}
\end{center}
\end{figure}

In Figure~\ref{fig-6} we show a J48 decision tree created using our data set. We introduced a pruning constraint for better understanding of its structure. The pruning constraint was introduced taking care of the balance between data description and performance, limiting the effect of the pruning constraint to 5 percentual points in F-measure. 

As Figure~\ref{fig-6} shows, the most relevant feature for this task is the authority of the second vertex, result that is consistent with the information gain analysis. In fact, a high value in authority for the second vertex (a.k.a. the candidate vertex) is enough evidence for real link prediction for 35,296 links over 2,993 false positives. Candidates with low authority scores need to be described using degree and transitivity. Finally, the characterization of the active is only marginally relevant for the problem, being considered for this task its degree. In fact, a low degree for the active user allows to detect real links for 80,346 cases over 33,628 false positives.

\begin{figure}[ht!]
\begin{center}
\includegraphics[width=0.7\columnwidth]{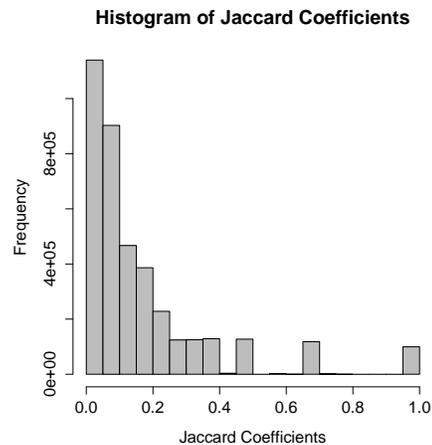}
\caption{Histogram for Jaccard Coefficients.}
\label{fig:jaccard_full}
\end{center}
\end{figure}

Now we will explore the use of locality thresholds for the selection of candidates. 
We calculated the Jaccard coefficient for each link in the graph stored until January 1st, that is to say, for the collection of 3,855,389 links created during 2013. The histogram for these coefficients is showed in Figure~\ref{fig:jaccard_full}.

As the histogram of Figure~\ref{fig:jaccard_full} shows, the main bin is in the origin, but a significant number of neighboors achieve high values for this measure. We will use three values for locality threshold using Jaccard: 0.1, 0.2 and 0.3. These values will allow us to retrieve a significant number of seeds for our vertex ranking algorithm. 

We will start our analysis from the collection of users that created at least one new friend during the observation period. This set of users has 76,848 vertices, and for each vertex in this set we retrieved the collection of neighboors with a Jaccard score over the threshold value. We retrieve 70,498, 61,056, and 34,929 seeds for Jaccard thresholds equals to 0.1, 0.2 and 0.3, respectively. Then, for each seed we retrieved its neighboors, which are considered as the friend candidates for each active user. Then, the list of links between active users and its candidates was checked in the graph. If a link was found in the graph stored until January 1st, we drop it from the collection. If a link was found in observation period (Jan 1st to 25), we labeled it as a real link. Finally, if a link was not found, we labeled it as a false link. 5,138, 1,875 and 810 real links were founded using our strategy, and a total amount of 235,348, 135,022 and 61,027 cases were labeled as false links, for threshold values equals to 0.1, 0.2 and 0.3, respectively. We balanced our data sets using resampling to avoid the analysis of biased results. 

We explore the feasibility of the ranking-based strategy by building classifiers for each labeled data set. Performance results are shown in Table~\ref{tab:ranking}.

\begin{table*}[h]
\begin{center}
\vspace{1cm}
{\setlength{\extrarowheight}{1.0pt}
\begin{tabular}{|l|l|c|c|c|c|c|c|c|c|}
\hline 
th = 0.1 & Class & TP Rate& FP Rate& Precision& Recall& F-Measure& MCC& ROC Area& PRC Area\\
\hline \hline
\multirow{3}{*}{Bayes} & real& 0.392& 0.064& 0.861& 0.392& 0.538& 0.391& 0.800& 0.799\\
\cline{2-10} 
& false& 0.936& 0.608& 0.606& 0.936& 0.736& 0.391& 0.800& 0.742\\
\cline{2-10} 
& Weighted Avg.& 0.664& 0.336& 0.733& 0.664& 0.637& 0.391& 0.800& 0.770\\
\hline \hline
\multirow{3}{*}{J48} & real& 0.920& 0.055& 0.943& 0.920& 0.931& 0.864& 0.975& 0.965\\
\cline{2-10} 
& false& 0.945& 0.080& 0.921& 0.945& 0.933& 0.864& 0.975& 0.978\\
\cline{2-10} 
& Weighted Avg.& 0.932& 0.068& 0.932& 0.932& 0.932& 0.864& 0.975& 0.971\\
\hline \hline
\multirow{3}{*}{Log} & real& 0.683& 0.168& 0.803& 0.683& 0.738& 0.521& 0.819& 0.813\\
\cline{2-10} 
& false& 0.832& 0.317& 0.724& 0.832& 0.774& 0.521& 0.819& 0.782\\
\cline{2-10} 
& Weighted Avg.& 0.758& 0.242& 0.764& 0.758& 0.756& 0.521& 0.819& 0.798\\
\hline \hline
th = 0.2 & Class & TP Rate& FP Rate& Precision& Recall& F-Measure& MCC& ROC Area& PRC Area\\
\hline \hline
\multirow{3}{*}{Bayes} & real& 0.559& 0.073& 0.884& 0.559& 0.685& 0.523& 0.837& 0.833\\
\cline{2-10} 
& false& 0.927& 0.441& 0.678& 0.927& 0.783& 0.523& 0.837& 0.787\\
\cline{2-10} 
& Weighted Avg.& 0.743& 0.257& 0.781& 0.743& 0.734& 0.523& 0.837& 0.810\\
\hline \hline
\multirow{3}{*}{J48} & real& 0.899& 0.064& 0.933& 0.899& 0.916& 0.836& 0.972& 0.967\\
\cline{2-10} 
& false& 0.936& 0.101& 0.903& 0.936& 0.919& 0.836& 0.972& 0.973\\
\cline{2-10} 
& Weighted Avg.& 0.917& 0.083& 0.918& 0.917& 0.917& 0.836& 0.972& 0.970\\
\hline \hline
\multirow{3}{*}{Log} & real& 0.742& 0.137& 0.844& 0.742& 0.790& 0.609& 0.844& 0.833\\
\cline{2-10} 
& false& 0.863& 0.258& 0.770& 0.863& 0.813& 0.609& 0.844& 0.802\\
\cline{2-10} 
& Weighted Avg.& 0.802& 0.198& 0.807& 0.802& 0.801& 0.609& 0.844& 0.818\\
\hline \hline
th = 0.3 & Class & TP Rate& FP Rate& Precision& Recall& F-Measure& MCC& ROC Area& PRC Area\\
\hline \hline
\multirow{3}{*}{Bayes} & real& 0.756& 0.097& 0.887& 0.756& 0.816& 0.666& 0.862& 0.854\\
\cline{2-10} 
& false& 0.903& 0.244& 0.787& 0.903& 0.841& 0.666& 0.862& 0.815\\
\cline{2-10} 
& Weighted Avg.& 0.829& 0.170& 0.837& 0.829& 0.829& 0.666& 0.862& 0.834\\
\hline \hline
\multirow{3}{*}{J48} & real& 0.919& 0.062& 0.937& 0.919& 0.928& 0.857& 0.975& 0.972\\
\cline{2-10} 
& false& 0.938& 0.081& 0.920& 0.938& 0.929& 0.857& 0.975& 0.975\\
\cline{2-10} 
& Weighted Avg.& 0.928& 0.072& 0.929& 0.928& 0.928& 0.857& 0.975& 0.973\\
\hline \hline
\multirow{3}{*}{Log} & real& 0.823& 0.102& 0.890& 0.823& 0.855& 0.723& 0.871& 0.864\\
\cline{2-10} 
& false& 0.898& 0.177& 0.835& 0.898& 0.865& 0.723& 0.871& 0.824\\
\cline{2-10} 
& Weighted Avg.& 0.860& 0.140& 0.862& 0.860& 0.860& 0.723& 0.871& 0.844\\
\hline 
\end{tabular}}
\end{center}
\caption{Performance measures for the real/false link classification problem using a locality threshold for the selection of candidates.}
\label{tab:ranking}
\end{table*}

As Table~\ref{tab:ranking} shows, the results achieved by our strategy outperforms the first approach. The use of locality threshold values impacts the performance by several percentual points, fact that illustrates the precision/recall tradeoff. A high threshold value limits the amount of seeds and accordingly, the amount of candidates is reduced. Thus, the precision of the method gets improvements but with low recall rates. On the other hand, the use of low threshold values allow to recover more seeds, but reducing the precision of the classifiers. 

\section{Conclusions}
\label{sec:conclusions}

We explored two approaches for the link prediction problem. 
A first approach models the problem as a real/false link classification problem. 
The second one uses locality thresholds to define a collection of seeds from where a list of candidates is generated. Our results show that the use of locality thresholds is effective in this problem, and can be succesfully combined with global measures to detect potential links. 

There are a number of practical issues involved with link prediction that needs to be addressed. 
First, the implementation of a ranking-based algorithm for link prediction requires the use of 
data structures that can efficiently deal with graphs. In our experience, the explosive growth of the collection of potential links is a hard task to be efficiently implemented using conventional data structures. Then an open question relies on the use of these algorithms on very large graphs.
In addition, how to efficiently update an index with locality measures is also an open question, considering that social networks are time evolving graphs.  

\bibliographystyle{plain}

\end{document}